\documentclass[aip,cha,reprint]{revtex4-1}
\usepackage{amsmath,amssymb,graphicx,color,bm,natbib,placeins,times,xspace,hyperref}

\begin{document} 

\title{Network-complement transitions, symmetries, and cluster synchronization}

\author{Takashi Nishikawa}
\affiliation{Department of Physics and Astronomy, Northwestern University, Evanston, IL 60208, USA}
\affiliation{Northwestern Institute on Complex Systems, Northwestern University, Evanston, IL 60208, USA}

\author{Adilson E. Motter} 
\affiliation{Department of Physics and Astronomy, Northwestern University, Evanston, IL 60208, USA}
\affiliation{Northwestern Institute on Complex Systems, Northwestern University, Evanston, IL 60208, USA}

\begin{abstract}
Synchronization in networks of coupled 
oscillators is known to be 
largely determined by
the spectral and symmetry properties of the 
interaction network.
Here we leverage this relation 
to
study a class of 
networks for which the threshold coupling strength for global synchronization is the lowest among all networks with the same number of nodes and links.
These networks, 
defined
as being \underline{u}niform, \underline{c}omplete, and \underline{m}ulti-partite (UCM), appear at each of an infinite sequence of network-complement transitions in a larger class of networks 
characterized by having
near-optimal
thresholds for global synchronization.
We show that the distinct symmetry structure of the UCM networks, which by design are optimized for global synchronizability, often  
leads to
formation of 
clusters
of synchronous 
oscillators, and that such states can coexist with the state of global synchronization.

\end{abstract}

\hfill \href{https://doi.org/10.1063/1.4960617}{\small Chaos {\bf 26}, 094818 (2016); DOI: 10.1063/1.4960617}

\maketitle

\begin{quotation}

The study of dynamical processes on networks has traditionally been centered on establishing relations between the collective behavior and structures in the network of existing interactions.\cite{Porter:2016}
Here we focus on the role played more explicitly by structures in the complement of the network, which can be interpreted as the network of absent interactions.
While the information provided by the network and its complement are mathematically equivalent, the interpretation of the results is easier and more natural in the complement. 
We focus on the class of networks with \underline{m}inimum possible size of the largest \underline{c}omponents in their \underline{c}omplement (MCC). We show that 1) MCC networks approach those that maximize global synchronizability (i.e., that require the smallest coupling strength for stable synchronization), 2) MCC networks exhibit an infinite sequence of transitions, each defining a UCM network characterized by being strictly optimal, and 3) the dynamics of UCM networks can be rich yet permits systematic identification of stability 
thresholds.
This work shows that the optimization of network structure for synchronization leads to significantly different results and is surprisingly more involved for undirected networks considered here when compared to the previously solved problem of directed networks\cite{Nishikawa:2010fk}. 

\end{quotation}

\section{Introduction}

In the study of networks there are two major classes of transitions that can be observed as an external parameter is varied: {\em structural transitions}, in which a sudden change occurs in the connectivity structure of the network, and {\em dynamical transitions}, in which the change occurs in the collective behavior of a network of coupled dynamical 
entities.
A primary example of a structural transition is the widely studied problem of network percolation,\cite{Dorogovtsev:2008,barrat2008dynamical} in which the relative size of the largest connected component of random networks exhibits a phase transition as the average number of links per node increases.
It has recently been discovered\cite{Achlioptas2009} that such transitions can be ``explosive,'' which was followed by 
a surge of
investigation on their exact nature,\cite{da-Costa-2010,Riordan-2011,DSouza:2015} and has highlighted the fact that even single-link modification can have large impact on global structural properties.\cite{Nagler2011}
An influential example of a dynamical transition is the sudden emergence of a synchronized subpopulation in Kuramoto's network of globally coupled phase oscillators\cite{kuramoto1984chemical} as the coupling strength increases, which was more recently extended to the case of more complex coupling structures.\cite{Watts:1999,Hong:2002,Ichinomiya:2004,Juan-G.-Restrepo:2006ak}
It has been shown that such synchronization transitions can also be explosive when correlation exists between the natural frequencies of the oscillators and the network structure.\cite{Gomez-Gardenes-2011,Ji-2013}
Here we consider transitions involving both structure and dynamics:  structural transitions in the class of MCC networks, which is defined through the impact their network structure has on a (different) dynamical transition---namely, the stability transition for global synchronization.
As the space of MCC networks is traversed by varying the total number of links in the network, structural transitions in the complement of the network is observed at special points that defines the UCM networks.

Optimizing the network structure for dynamics is a widely studied problem in the context of networks and complex systems,
particularly in the context of maximizing global
synchronizability.\cite{PhysRevE.81.025202,PhysRevLett.95.188701,Nishikawa:2006fk,Nishikawa:2006kx,Wang:2007kx,6561538,Nishikawa:2010fk,Motter2004af}
Recent studies on network synchronization dynamics\cite{Tang:2014,Pecora:2015,Pikovsky:2015} have also encompassed various other forms of synchronization, such as cluster synchronization,\cite{Zhou:2006,Dahms:2012,Pecora:2014zr} remote synchronization,\cite{Bergner:2012,Nicosia:2013,Gambuzza:2013} relay synchronization,\cite{Fischer:2006} and chimera states.\cite{Kuramoto:2002,Abrams:2004}
Here we show that the UCM networks, despite being designed to optimize global synchronization, often exhibit cluster synchronization.
This counter-intuitive effect can be attributed to their highly symmetric structure, which is a byproduct of the optimization that also allows us to develop systematic stability analysis for cluster synchronization.

\section{Stability of global synchronization \label{sec:global-sync}}

The dynamical model we use for a network of coupled oscillators is the Pecora-Carroll model\cite{Pecora:1998zp} with the following notation.
The governing equation reads
\begin{equation}\label{eqn:syst}
\dot{\mathbf{x}}_i = \mathbf{F}(\mathbf{x}_i) + \varepsilon \sum_{j=1}^n A_{ij} [\mathbf{H}(\mathbf{x}_j) - \mathbf{H}(\mathbf{x}_i)], \quad i = 1,\ldots,n.
\end{equation}
We assume that the dynamics of an isolated oscillator, $\dot{\mathbf{x}}_i = \mathbf{F}(\mathbf{x}_i)$, is chaotic and identical for all oscillators.
We also assume that the network encoded by the adjacency matrix $A = (A_{ij})_{1 \le i,j \le n}$ is undirected (i.e., $A_{ij} = A_{ji}$), unweighted (i.e., $A_{ij} = 1$), and connected.
Using the corresponding Laplacian matrix $L$, defined by
\begin{equation}
L_{ij} = \begin{cases}
-A_{ij} &\text{if $i \neq j$,}\\
\sum_{k\neq i} A_{ik} &\text{if $i=j$,}
\end{cases}
\end{equation}
system~\eqref{eqn:syst} can be written as
\begin{equation}\label{eqn:syst-lap}
\dot{\mathbf{x}}_i = \mathbf{F}(\mathbf{x}_i) - \varepsilon \sum_{j=1}^n L_{ij} \mathbf{H}(\mathbf{x}_j).
\end{equation}
For any given dynamics of an isolated oscillator $\mathbf{x}_\text{GS} = \mathbf{x}_\text{GS}(t)$ (satisfying $\dot{\mathbf{x}}_\text{GS} = \mathbf{F}(\mathbf{x}_\text{GS})$), there is a corresponding solution of Eq.~\eqref{eqn:syst} describing a state of {\em global synchronization} (GS), given by $\mathbf{x}_i(t) = \mathbf{x}_\text{GS}(t)$ for all $i$ and all $t$.
The stability of this solution can be analyzed using the master stability function approach~\cite{Pecora:1998zp}, which is based on the linearization of Eq.~\eqref{eqn:syst} around the globally synchronous state and the transformation of the state variables to eigenvector coordinates of the Laplacian matrix $L$.  The result of this analysis is a master stability function (MSF), which is denoted by $\Lambda_\text{GS}(\alpha)$ and defined to be the maximum Lyapunov exponent of
\begin{equation}\label{eqn:mse}
\delta\dot{\mathbf{x}} = [D\mathbf{F}(\mathbf{x}_\text{GS}) - \alpha D\mathbf{H}(\mathbf{x}_\text{GS})] \delta\mathbf{x},
\end{equation}
where $\alpha$ is an auxiliary parameter.
The condition for the stability of the globally synchronous state is 
\begin{equation}
\max_{2\le j \le n} \Lambda_\text{GS}(\varepsilon\lambda_j) < 0, 
\end{equation}
where $0 = \lambda_1 < \lambda_2 \le \cdots \le \lambda_n$ are the eigenvalues of $L$. 
Note that $\lambda_1 = 0$ because the Laplacian is a zero row sum matrix, $\lambda_j$ are all real because $L$ is a symmetric matrix for undirected networks, and $\lambda_2>0$ because the network is assumed to be connected.
Since the oscillators are chaotic, GS is unstable when there is no coupling between them (i.e., when $\varepsilon=0$), implying $\Lambda_\text{GS}(0) > 0$.
For any MSF that is negative on a semi-infinite interval (i.e., $\Lambda_\text{GS}(\alpha) < 0$ on $(\alpha_1,\infty)$ with $\alpha_1>0$), the threshold coupling strength---the minimum value of $\varepsilon$ for which GS is stable---is given by $\alpha_1/\lambda_2 \equiv \varepsilon^\text{th}_1$.
For an MSF that is negative on a finite interval $(\alpha_1,\alpha_2)$ with $0<\alpha_1<\alpha_2$, there can be networks that are not synchronizable at all (i.e., have no value of $\varepsilon$ that allows $\alpha_1<\varepsilon\lambda_2 \le \varepsilon\lambda_n < \alpha_n$ to be satisfied).
In that case, the value $\varepsilon^\text{th}_1$ is the coupling threshold for GS for all synchronizable networks.
Thus, the larger the value of $\lambda_2$, the lower the threshold for the stability of GS.
The eigenvalue $\lambda_2$ is also known to be important for other processes, such as diffusion dynamics,\cite{chung1997spectral,Motter:2005ub} consensus protocol,\cite{4140748,4700861} and Turing instability in activator-inhibitor systems.\cite{Nakao:2010fk}

\section{Network-complement transitions}

The problem of identifying and characterizing the networks maximizing $\lambda_2$ (and thus minimizing $\varepsilon^\text{th}_1$) for a given number of nodes and links is a challenging problem that has been a topic of investigation in the mathematics communities.\cite{maas1987transportation,Alon:1986uq,Friedman:1989:SER:73007.73063,Lubotzky:1988fk}
As we see below, in some cases this problem can be completely solved.
Indeed, if $n = k\ell$ with integers $k$ and $\ell$ and the number of links is $m = \frac{1}{2}k^2\ell(\ell-1)$, it can be shown that the maximum possible value $\lambda_2 = n - k$ is achieved only by the {\em UCM network}, defined as the network having $\ell$ groups of $k$ oscillators (i.e., \underline{u}niform group size) in which the groups are fully connected to each other (i.e., \underline{c}omplete), but have no internal links within any group (i.e., \underline{m}ulti-partite).
For other combinations of $n$ and $m$, a class of networks that share a unique property with the UCM networks can be defined and shown\cite{sensitivity} to have maximal or near-maximal $\lambda_2$ values for a wide range of the parameters $n$ and $m$.
We refer to such networks as {\em MCC networks}, since they are defined to be the networks that have the \underline{m}inimum possible size of the largest \underline{c}omponents in their graph \underline{c}omplement among all networks with the same $n$ and $m$.
Here, {\em components} of a given network refer to the connected components of the network, and 
the {\em complement} of a given network with adjacency matrix $A$ is defined as the network with the adjacency matrix $A^c$ given by $A^c_{ij} = 1$ if $A_{ij} = 0$ with $i \neq j$, $A^c_{ij} = 0$ if $A_{ij} = 1$ with $i \neq j$, and $A_{ii} = 0$.
It follows that the UCM networks, whose complement consists of $\ell$ fully connected components of size $k$, are special cases of MCC networks.

\begin{figure}
\vspace{4mm}
\begin{center}
\includegraphics[width=\columnwidth]{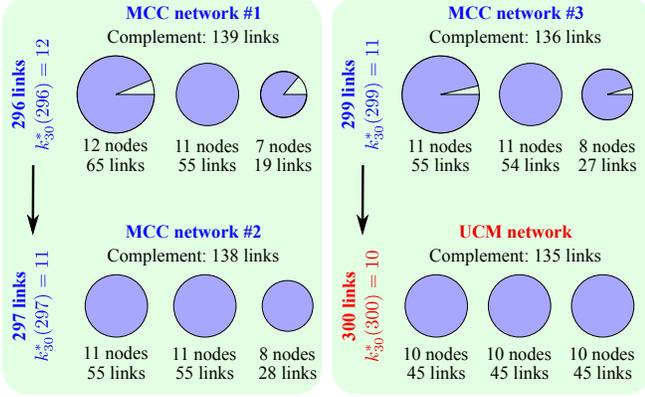}
\end{center}
\vspace{-5mm}
\caption{\label{ucm-mcc}
Examples of network-complement transitions for MCC networks.
For each network we visualize its complement, using circles to indicate the connected components and the shaded area to indicate the density of links in these components.
(Left) After a single link is added to MCC network \#1, it becomes possible to rearrange links to decrease the largest component size in the complement, $k_n^*(m)$, and obtain MCC network \#2.
(Right) By adding a link to MCC network \#3, it also becomes possible to rearrange links not only to decrease $k_n^*(m)$, but also to obtain a UCM network in this case.}
\end{figure}

We now study the structural properties of the complement of the UCM and MCC networks as a function of the number of links.
Let $k_n^*(m)$ denote the size of the largest components in the complement of the MCC network with $n$ nodes and $m$ links.
This integer-valued function can be expressed as
\begin{equation}\label{eqn:kn}
k^*_n(m) = \lceil C_{n,m} \cdot n \rceil, \quad 
C_{n,m} \equiv \frac{\ell + \sqrt{\ell^2 - \frac{2m}{n^2}\ell(\ell+1)}}{\ell(\ell+1)},
\end{equation}
where we denote by $\lceil x \rceil$ the smallest integer greater than or equal to $x$, and we define $\ell$ to be the (unique) integer satisfying \
$m^*_{\ell,n} \le m < m^*_{\ell+1,n}$ with $m^*_{\ell,n} \equiv \frac{n^2(\ell-1)}{2\ell}$.
As $m$ increases from $m = n-1$ (minimum required for a connected network) to $m = \frac{1}{2}n(n-1)$ (complete graph) while keeping $n$ fixed, function $k^*_n(m)$ decreases from $n$ to $1$, jumping down by one at various points, which are non-uniformly distributed over the range of $m$ values.
These jumps are associated with the addition of a link (i.e., deletion of a link in the complement) that for the first time allows for a smaller component size in the complement through global rearrangement of the links (see Fig.~\ref{ucm-mcc} for examples). 
Note that some jumps occur when the network becomes UCM (e.g., at $m=300$ in Fig.~\ref{ucm-mcc}, right column), while others occur without the appearance of a UCM network (e.g., at $m=297$ in Fig.~\ref{ucm-mcc}, left column).

\begin{figure}
\begin{center}
\includegraphics[width=\columnwidth]{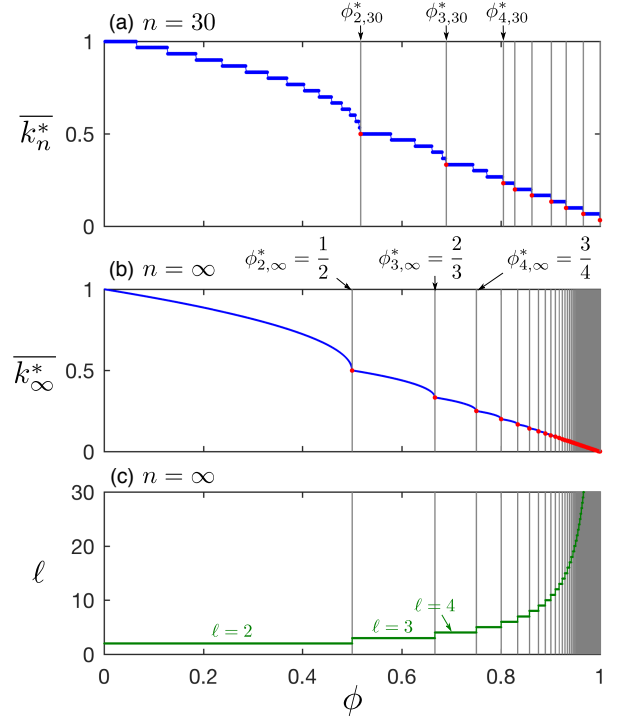}
\end{center}
\vspace{-7mm}
\caption{\label{transitions_fig}
Infinite sequence of transitions in the complement of MCC networks.
(a) Relative size of the largest connected components $\overline{k^*_n}$ in the complement of MCC networks with $n=30$ nodes as a function of link density $\phi$.
Vertical lines and red dots indicate the points 
$\phi = \phi^*_{\ell,30}$, 
where $\overline{k^*_n}$ jumps down by one.
(b) $\overline{k^*_\infty}$ given by Eq.~\eqref{eqn:n-inf} for the limit of large networks, $n \to \infty$.
(c) $\ell$ in the limit $n \to \infty$, which can be interpreted as the number of components in the complement.
The infinite sequence of non-differentiable points of $\overline{k^*_\infty}$ (vertical lines and red dots in (b)) indicates the transitions at 
$\phi = \phi^*_{\ell,\infty}$, 
$\ell=2,3,\ldots$, each of which is accompanied by a discrete change in $\ell$ shown in (c).}
\end{figure}

In order to consider the limit of large network size, we normalize the maximum component size in the complement, $k^*_n(m)$, by its upper bound, $n$, and express it as a function of the link density, $\phi \equiv \frac{2m}{n(n-1)}$, to obtain the relative component size $\overline{k^*_n}(\phi) \equiv \frac{1}{n} k_n^*\left(\frac{1}{2}\phi n(n-1)\right)$.
Figure~\ref{transitions_fig}(a) shows an example of $\overline{k^*_n}(\phi)$ for MCC networks of size $n=30$.
We observe that $\overline{k^*_n}(\frac{2}{n}) = 1$ and $\overline{k^*_n}(1) = \frac{1}{n}$, with the jumps of size $\frac{1}{n}$ at $n-1$ intermediate points. 
Defining 
$\phi^*_{\ell,n} \equiv \frac{2m^*_{\ell,n}}{n(n-1)}$
and taking the limit $m,n \to \infty$ with fixed $\phi$ in Eq.~\eqref{eqn:kn}, we obtain
\begin{equation}\label{eqn:n-inf}
\overline{k^*_\infty}(\phi) 
= \frac{\ell + \sqrt{\ell^2 - \phi\ell(\ell+1)}}{\ell(\ell+1)},
\end{equation}
where $\ell$ is the (unique) integer satisfying 
$\phi^*_{\ell,\infty} \le \phi < \phi^*_{\ell+1,\infty}$ with $\phi^*_{\ell,\infty} \equiv \lim_{n \to \infty} \phi^*_{\ell,n} = \frac{\ell-1}{\ell}$.
The function $\overline{k^*_\infty}(\phi)$ given by Eq.~\eqref{eqn:n-inf} and shown in Fig.~\ref{transitions_fig}(b) has an infinite number of non-differentiable points at 
$\phi^*_{\ell,\infty}$, $\ell=2,3,{\ldots}\xspace\,.$
Note that the integer $\ell$ can be interpreted as the number of connected components in the complement in the large-network limit.
The transition points 
$\phi^*_{\ell,\infty}$, 
at which $\ell$ makes discrete jumps, correspond to UCM networks with $\ell$ groups of equal size.
Details of the proofs and derivations on the properties of the MCC and UCM networks can be found in Ref.~\onlinecite{sensitivity}.

\section{Patterns of Cluster Synchronization}

We now go beyond the GS stability threshold $\varepsilon^\text{th}_1$ and study the emergence of synchronization patterns in the UCM networks.
These networks are optimized for GS, and we now study their partial synchronization properties.
As we show next, the group structure of the UCM network allows us to carry out the stability analysis explicitly for cluster synchronization.
To reflect this group structure, let us re-index the nodes and denote the state vector for the $i$-th node in the $h$-th group by $\mathbf{x}_i^{(h)}$.
Using this notation and exploiting the group structure of the UCM networks, Eq.~\eqref{eqn:syst-lap} becomes
\begin{equation}\label{eqn:syst-group}
\dot{\mathbf{x}}_i^{(h)} = \mathbf{F}(\mathbf{x}_i^{(h)}) 
- \varepsilon k(\ell-1) \mathbf{H}(\mathbf{x}_i^{(h)})
+ \varepsilon \sum_{\substack{h'=1\\ h'\neq h}}^\ell \sum_{j=1}^k \mathbf{H}(\mathbf{x}_j^{(h')}).
\end{equation}
Now consider a state of {\em cluster synchronization} (CS) given by $\mathbf{x}_i^{(h)}(t) = \mathbf{x}_\text{CS}^{(h)}(t)$ for all $i=1,\ldots,k$ and $h = 1,\ldots,\ell$.
Substituting this into Eq.~\eqref{eqn:syst-group}, we obtain an equation of the same form as Eq.~\eqref{eqn:syst-group} but for the network of $\ell$ nodes, each representing a group of $k$ (synchronized) oscillators:
\begin{equation}\label{eqn:syst-group-sync}
\dot{\mathbf{x}}_\text{CS}^{(h)} = \mathbf{F}(\mathbf{x}_\text{CS}^{(h)})
- \varepsilon k \sum_{h'=1}^\ell \widetilde{L}_{hh'} \mathbf{H}(\mathbf{x}_\text{CS}^{(h')}),
\end{equation}
where $\widetilde{L} = (\widetilde{L}_{hh'})_{1 \le h,h' \le \ell}$ is the Laplacian matrix of the fully connected network with $\ell$ nodes.
This is the equation that must be satisfied by the CS state.
The special case, $\mathbf{x}_\text{CS}^{(h)} = \mathbf{x}_\text{GS}(t)$ for all $h$, which satisfies Eq.~\eqref{eqn:syst-group-sync}, describes the synchronization between the groups (in addition to the synchronization of oscillators within each group) and thus corresponds to the globally synchronous state discussed in Sec.~\ref{sec:global-sync}.
We can apply the same MSF analysis, since 
Eq.~\eqref{eqn:syst-group-sync} is just a slight modification of Eq.~\eqref{eqn:syst-lap}; we obtain Eq.~\eqref{eqn:syst-group-sync} from Eq.~\eqref{eqn:syst-lap} by replacing $\varepsilon$ with $\varepsilon k$ and redefining $\widetilde{L}_{hh'}$ to represent the all-to-all coupling structure.
It follows that the synchronization between the groups is determined by $\Lambda_\text{GS}(\varepsilon k \ell) = \Lambda_\text{GS}(\varepsilon n)$, since the only nontrivial Laplacian eigenvalue of the size-$\ell$ fully connected network is $\ell$.
The Laplacian eigenvalues of a UCM network can be calculated and are $0$, $n-k$, and $n$, with multiplicity $1$, $\ell(k-1)$, and $\ell-1$, respectively.
Among these the eigenvalue $n$ corresponds to $\Lambda_\text{GS}(\varepsilon n)$, thus associating this eigenvalue with the modes of perturbation that can destroy GS, but not the synchronization within each group.

To analyze the synchronization stability of individual groups, we linearize Eq.~\eqref{eqn:syst-group} around the CS state, which leads to the following variational equation:
\begin{equation}\label{eqn:syst-group-var}
\begin{split}
\delta\dot{\mathbf{x}}_i^{(h)} = 
[D\mathbf{F}(\mathbf{x}_\text{CS}^{(h)}) 
- \varepsilon k(\ell-1) D\mathbf{H}(\mathbf{x}_\text{CS}^{(h)})] \delta\mathbf{x}_i^{(h)}\\
- \varepsilon \sum_{\substack{h'=1\\ h'\neq h}}^\ell D\mathbf{H}(\mathbf{x}_\text{CS}^{(h')}) \sum_{j=1}^k \delta\mathbf{x}_j^{(h')}.
\end{split}
\end{equation}
If we define $S^{(h)} \equiv \sum_{j=1}^k \delta\mathbf{x}_j^{(h)}$ and sum Eq.~\eqref{eqn:syst-group-var} over $i$, we obtain
\begin{equation}\label{eqn:syst-group-var-sum}
\begin{split}
\dot{S}^{(h)} = 
[D\mathbf{F}(\mathbf{x}_\text{CS}^{(h)}) 
- \varepsilon k(\ell-1) D\mathbf{H}(\mathbf{x}_\text{CS}^{(h')})] S^{(h)}\\
- \varepsilon \sum_{\substack{h'=1\\ h'\neq h}}^\ell D\mathbf{H}(\mathbf{x}_\text{CS}^{(h')}) S^{(h')}.
\end{split}
\end{equation}
Now consider those perturbations that do not affect the synchronization of any group, i.e., those for which $S^{(h)}(0) = \sum_{j=1}^k \delta\mathbf{x}_i^{(h)}(0) = 0$ for all $h$ and all $i$.  
From Eq.~\eqref{eqn:syst-group-var-sum}, we see that $S^{(h)}(t) = \sum_{j=1}^k \delta\mathbf{x}_j^{(h)}$ remains zero at all times, and hence that the second term in Eq.~\eqref{eqn:syst-group-var} vanishes.
This collapses the $k$ equations for group $h$ into a single equation:
\begin{equation}\label{eqn:cs-mse}
\delta\dot{\mathbf{x}}^{(h)} = 
[D\mathbf{F}(\mathbf{x}_\text{CS}^{(h)}) 
- \varepsilon k(\ell-1) D\mathbf{H}(\mathbf{x}_\text{CS}^{(h)})] \delta\mathbf{x}^{(h)}.
\end{equation}
Thus, the stability against all perturbations that affect the synchronization of group $h$ is determined by the same equation, while this stability can be different for different groups through the dependence of Eq.~\eqref{eqn:cs-mse} on $\mathbf{x}_\text{CS}^{(h)}$.
The lack of dependence on $h' \neq h$ (i.e., on the states of other groups) in Eq.~\eqref{eqn:cs-mse} indicates that we have completely decoupled the problem of synchronization stability for individual groups.
Altogether, we have shown that the stability of CS can be analyzed by (i)~solving the system of $\ell$ coupled oscillators in Eq.~\eqref{eqn:syst-group-sync} to obtain $\mathbf{x}_\text{CS}^{(h)}$, and then (ii) solving Eq.~\eqref{eqn:cs-mse} for each $h=1,\ldots,\ell$.
We remark that the result above can also be derived using a general method based on the irreducible representations of the symmetry group\cite{Pecora:2014zr}.
Here, we were able to derive the results in a simpler and clearer fashion by explicitly using the special structure of the UCM networks, namely the groups with identical size and connectivity patterns.
This, however, suggests that there may be other classes of networks for which a similar simplification of the stability analysis is possible (i.e., an equation similar to Eq.~\eqref{eqn:cs-mse} can be derived).  While determining and classifying all possible CS solutions and their stability from the network's symmetry group is challenging, there have been a number of studies tackling this problem by focusing on specific classes of networks or on specific aspects of the problem\cite{Benoit:1996,Belykh:2000,Pogromsky:2002,Belykh:2003,Pogromsky:2008,Lin:2016,Pecora:2014zr,Sorrentino:2016}.

\begin{figure}
\begin{center}
\includegraphics[width=\columnwidth]{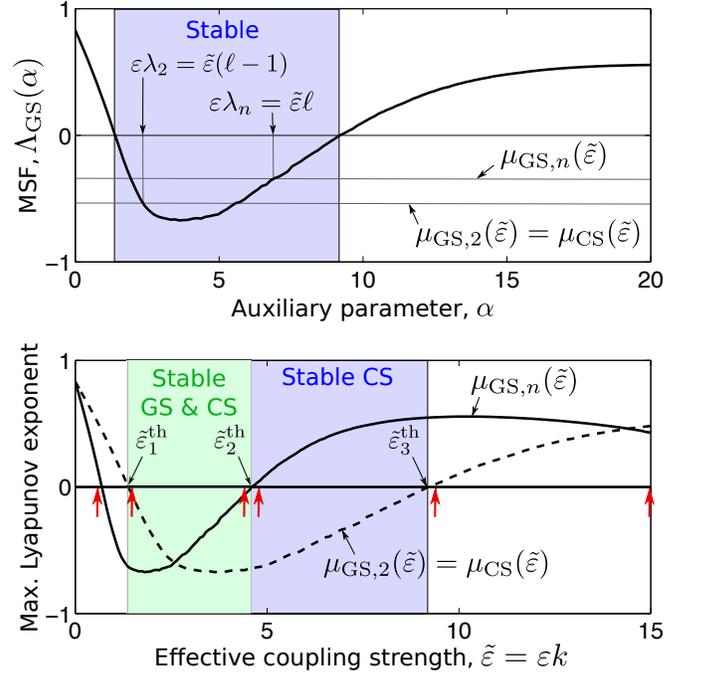}
\end{center}
\vspace{-5mm}
\caption{\label{fig:msf_plot}
Stability condition for GS and CS states in UCM networks of coupled Lorenz oscillators with $\ell=2$ groups of arbitrary size $k$.
(a) MSF as a function of the auxiliary parameter $\alpha$.
The condition for the GS state to be stable is to have both $\varepsilon\lambda_2$ and $\varepsilon\lambda_n$ in the blue-shaded region.
(b) The Lyapunov exponents $\mu_{\text{GS},2}(\tilde{\varepsilon}) = \mu_\text{CS}(\tilde{\varepsilon})$ and $\mu_{\text{GS},n}(\tilde{\varepsilon})$ plotted as a function of the effective coupling strength $\tilde{\varepsilon}$.
The red arrows indicate the values of $\tilde{\varepsilon}$ used in the full-network simulations shown in Figs.~\ref{fig:direct-sim-lorenz}--\ref{fig:direct-sim-lorenz-3}.}
\end{figure}

As an example, we consider a network of coupled Lorenz oscillators governed by Eq.~\eqref{eqn:syst} with
\begin{equation}
\mathbf{F}(\mathbf{x}) = \begin{pmatrix}
\sigma (y - x)\\
x(\rho - z) - y\\
xy - \beta z
\end{pmatrix}, \quad
\mathbf{H}(\mathbf{x}) = \begin{pmatrix} 0\\ 0\\ z \end{pmatrix}, \quad
\mathbf{x} = \begin{pmatrix} x\\ y\\ z \end{pmatrix},
\end{equation}
and the standard parameter values, $\sigma = 10$, $\rho = 28$, and $\beta = 2$, for which an isolated oscillator is chaotic.\cite{Lorenz:1963}
Our choice of $\mathbf{H}(\mathbf{x})$ represents the coupling among the oscillators through their $z$-components.
In terms of $x$-, $y$-, and $z$-components, Eq.~\eqref{eqn:syst-lap} reads
\begin{equation}\label{eqn:syst-lap-lorenz}
\begin{split}
\dot{x}_i &= \sigma (y_i - x_i),\\
\dot{y}_i &= x_i(\rho - z_i) - y_i,\\
\dot{z}_i &= x_i y_i - \beta z_i - \varepsilon \sum_{j=1}^n L_{ij} z_j.
\end{split}
\end{equation}
Note that this coupled system has a unique symmetry: it is invariant under the transformation $\{ x_i \to -x_i, \,\, y_i \to -y_i\}$ for any given $i$.
This is due to the symmetry of the individual Lorenz system about the $z$-axis in phase space, combined with the diffusive nature of the coupling (and thus the invariance holds for an arbitrary network structure).
Because of the symmetry, there are $2^{n-1}$ distinct anti-phase synchronous states of the system, each of which can be encoded by a length-$n$ binary sequence $a = (a_1, a_2, \ldots, a_n)$ with $a_i = \pm 1$.
(The number of distinct states is $2^{n-1}$ rather than $2^n$ because the two sequences related by the flipping of the sign of all $a_i$ both correspond to the same state.)
To construct these states, let $\mathbf{x}_\text{GS} = \mathbf{x}_\text{GS}(t) = (x_\text{GS}(t), y_\text{GS}(t), z_\text{GS}(t))^T$ be the chaotic trajectory of the isolated oscillator (thus satisfying $\dot{\mathbf{x}}_\text{GS} = \mathbf{F}(\mathbf{x}_\text{GS})$).
For each binary pattern $a$, the state given by $x_i(t) = a_i x_\text{GS}(t)$, $y_i(t) = a_i y_\text{GS}(t)$, $z_i(t) = z_\text{GS}(t)$ for each $i$ can be readily checked to be a valid solution of Eq.~\eqref{eqn:syst-lap-lorenz}.
Thus, these states, including the globally synchronous state and the cluster synchronous states of UCM networks, are guaranteed to exist, but their stability depends on the network structure and the coupling strength.

Using the analysis from Sec.~\ref{sec:global-sync}, the stability of GS for the coupled Lorenz systems in Eq.~\eqref{eqn:syst-lap-lorenz} is determined by $\Lambda_\text{GS}(\alpha)$, which is shown in Fig.~\ref{fig:msf_plot}(a) and is negative on the finite interval $(\alpha_1,\alpha_2)$, where $\alpha_1 \approx 1.39$ and $\alpha_2 \approx 9.20$.
Since the Laplacian eigenvalues of a UCM network are $0$, $k(\ell-1)$, and $k\ell$, the stability condition is that both $\mu_{\text{GS},2}(\tilde{\varepsilon}) \equiv \Lambda_\text{GS}(\varepsilon k(\ell-1)) < 0$ and $\mu_{\text{GS},n}(\tilde{\varepsilon}) \equiv \Lambda_\text{GS}(\varepsilon k\ell) < 0$ hold.
Figure~\ref{fig:msf_plot}(a) illustrates this situation.
Note that we have defined the effective coupling strength $\tilde{\varepsilon} \equiv \varepsilon k$, that the quantities $\mu_{\text{GS},2}(\tilde{\varepsilon})$ and $\mu_{\text{GS},n}(\tilde{\varepsilon})$ are functions of  $\tilde{\varepsilon}$ for a given $\ell$, and hence that the stability condition depends on $k$ only through $\tilde{\varepsilon}$.
Thus, the stability of GS is lost
when $\mu_{\text{GS},2}(\tilde{\varepsilon})=0$ at
$\tilde{\varepsilon} = \tilde{\varepsilon}^\text{th}_1 \equiv \frac{\alpha_1}{\ell-1}$.
Since the function $\Lambda_\text{GS}(\alpha)$ has a second zero crossing, there is a second threshold, $\tilde{\varepsilon}^\text{th}_2 \equiv \frac{\alpha_2}{\ell}$, at which $\mu_{\text{GS},n}(\tilde{\varepsilon})=0$.
This is when the GS state loses stability.
We note that, if the MSF were negative on an semi-infinite interval, such loss of GS stability would not occur.

\begin{figure*}
\begin{center}
\includegraphics[width=\textwidth]{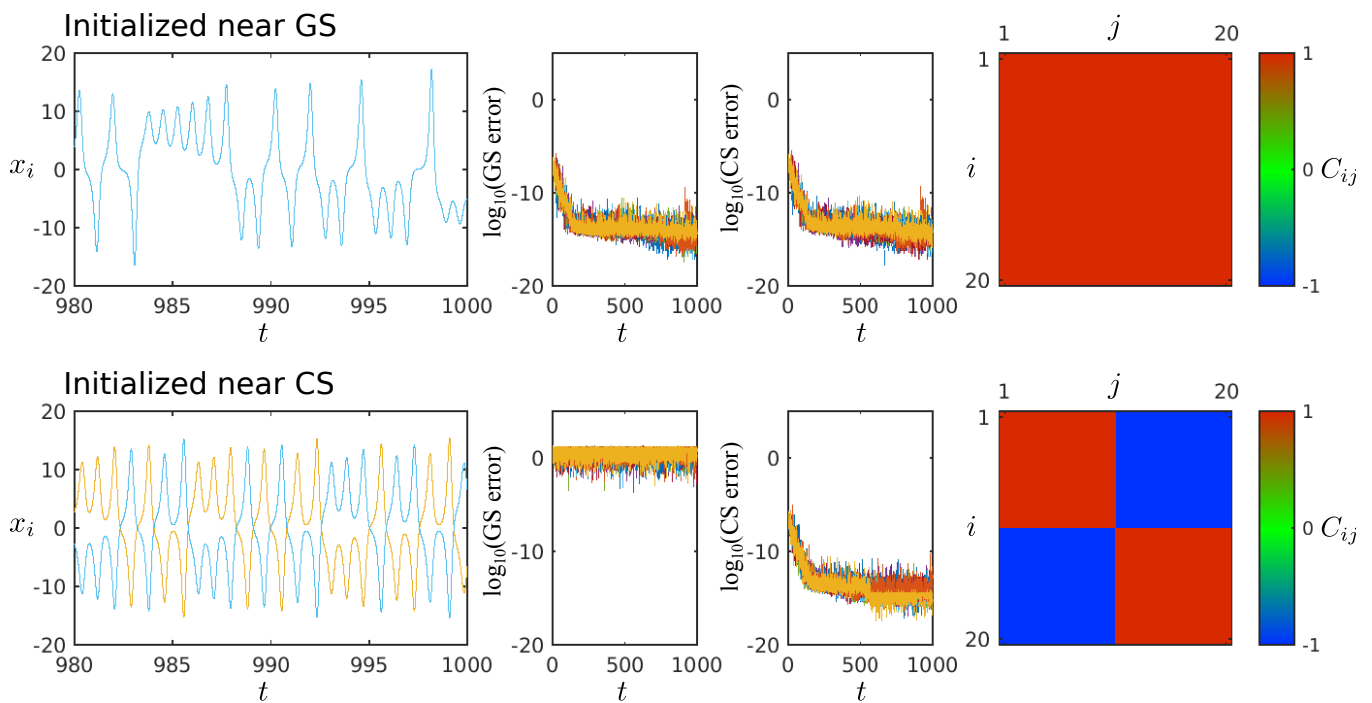}
\end{center}
\caption{\label{fig:direct-sim-lorenz}
Synchronization dynamics in UCM networks of coupled Lorenz oscillators with $\ell=2$ groups of size $k=10$ for various effective coupling strength $\tilde{\varepsilon}$.
(a) Ten independent simulations of the full network for $\tilde{\varepsilon} = 0.6$, each initialized at a distance of $10^{-6}$ from the anti-phase CS state.
(b) Simulation results for $\tilde{\varepsilon} = 1.5$.
In each row, the leftmost plot shows the $x$-component of all oscillators, $i=1,\ldots,k\ell=20$, for a single trajectory initialized as indicated.
The next two plots show the synchronization error measured by the standard deviation of $x_i$ over all $i$ for GS and the sum of within-group standard deviations for CS.
Ten separate trajectories are shown.
The last plot shows the matrix $C = (C_{ij})$, where $C_{ij}$ is the Pearson correlation coefficient of the $x$-component of oscillator $i$ and $j$ over $500 \le t \le 1000$.}
\end{figure*}

We now consider the cluster synchronous states of UCM networks.
First we note that a subset of the $2^n$ symmetry-based anti-phase synchronous states reflect the group structure of UCM networks and correspond to CS, whose stability we analyzed above.
We can see this in Eq.~\eqref{eqn:syst-group-sync}, which for Lorenz oscillators reads
\begin{equation}\label{eqn:syst-group-sync-lorenz}
\begin{split}
\dot{x}_\text{CS}^{(h)} &= \sigma (y_\text{CS}^{(h)} - x_\text{CS}^{(h)}),\\
\dot{y}_\text{CS}^{(h)} &= x_\text{CS}^{(h)} (\rho - z_\text{CS}^{(h)}) - y_\text{CS}^{(h)},\\
\dot{z}_\text{CS}^{(h)} &= x_\text{CS}^{(h)} y_\text{CS}^{(h)} - \beta z_\text{CS}^{(h)} - \varepsilon \sum_{h'=1}^\ell \widetilde{L}_{hh'} z_\text{CS}^{(h')}.
\end{split}
\end{equation}
For each binary pattern $a$ of length $\ell$, the state given by $x_\text{CS}^{(h)}(t) = a_h x_\text{GS}(t)$, $y_\text{CS}^{(h)}(t) = a_h y_\text{GS}(t)$, and $z_\text{CS}^{(h)}(t) = z_\text{GS}(t)$ for each $h$ gives a cluster synchronous state.
There are $2^{\ell-1}$ distinct states of this type.
Again, due to the symmetry, it can be shown for all $h$ that Eq.~\eqref{eqn:cs-mse} becomes identical to Eq.~\eqref{eqn:mse} with $\alpha=\varepsilon k(\ell-1)$, collapsing $\ell$ equations into one.
Thus, the stability condition for all these states is $\mu_\text{CS}(\tilde{\varepsilon}) \equiv \Lambda_\text{GS}(\varepsilon k(\ell-1)) = \mu_{\text{GS},2}(\tilde{\varepsilon}) < 0$.
Thus, for the system in Eq.~\eqref{eqn:syst-lap-lorenz} the CS states become stable at $\tilde{\varepsilon}^\text{th}_1$, which is identical to the stability threshold for the GS state.

Consider for simplicity the case of $\ell=2$, for which there are only two patterns: $a = (1,1)$ (corresponding to GS) and $a = (1,-1)$ (corresponding to anti-phase CS).
The functions $\mu_{\text{GS},2}(\tilde{\varepsilon}) = \mu_\text{CS}(\tilde{\varepsilon})$ and $\mu_{\text{GS},n}(\tilde{\varepsilon})$ are shown in Fig.~\ref{fig:msf_plot}.
Figure~\ref{fig:direct-sim-lorenz} shows the simulated dynamics of the full network for various values of $\tilde{\varepsilon}$ with $k=10$.
Below the first threshold $\tilde{\varepsilon}^\text{th}_1 \approx 1.39$, both the globally synchronous state and the cluster synchronous state are unstable, resulting in near-zero correlation between all pairs of oscillators (Fig.~\ref{fig:direct-sim-lorenz}(a)).
Above $\tilde{\varepsilon}^\text{th}_1$, we observe bi-stability: both types of synchronous states coexist and are stable (Fig.~\ref{fig:direct-sim-lorenz}(b)).
At $\tilde{\varepsilon}^\text{th}_2 \approx 4.60$, where $\Lambda_\text{GS}(\varepsilon k\ell) = 0$, GS loses its stability.
Increasing $\tilde{\varepsilon}$ further, we observe that the CS state become unstable at $\tilde{\varepsilon}^\text{th}_3 \approx 9.20$, where 
$\Lambda_\text{GS}(\varepsilon k(\ell-1)) = 0$.

\begin{figure*}
\begin{center}
\includegraphics[width=\textwidth]{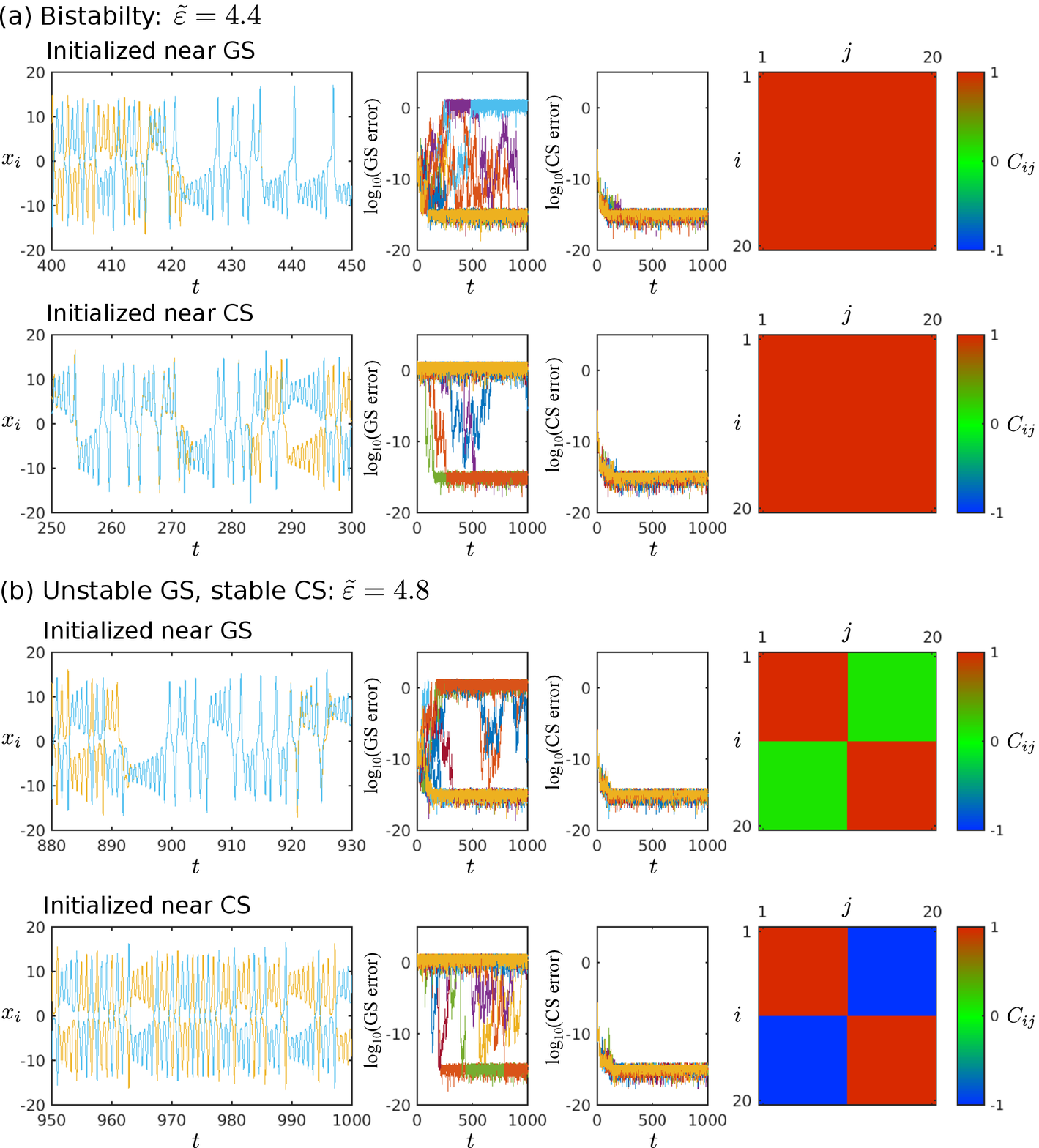}
\end{center}
\caption{\label{fig:direct-sim-lorenz-2}
Synchronization dynamics in UCM networks of coupled Lorenz oscillators with $\ell=2$ groups of size $k=10$ for $\tilde{\varepsilon}=4.4$ and $\tilde{\varepsilon}=4.8$. Detail description of individual plots is the same as in Fig.~\ref{fig:direct-sim-lorenz}.}
\end{figure*}

\begin{figure*}
\begin{center}
\includegraphics[width=\textwidth]{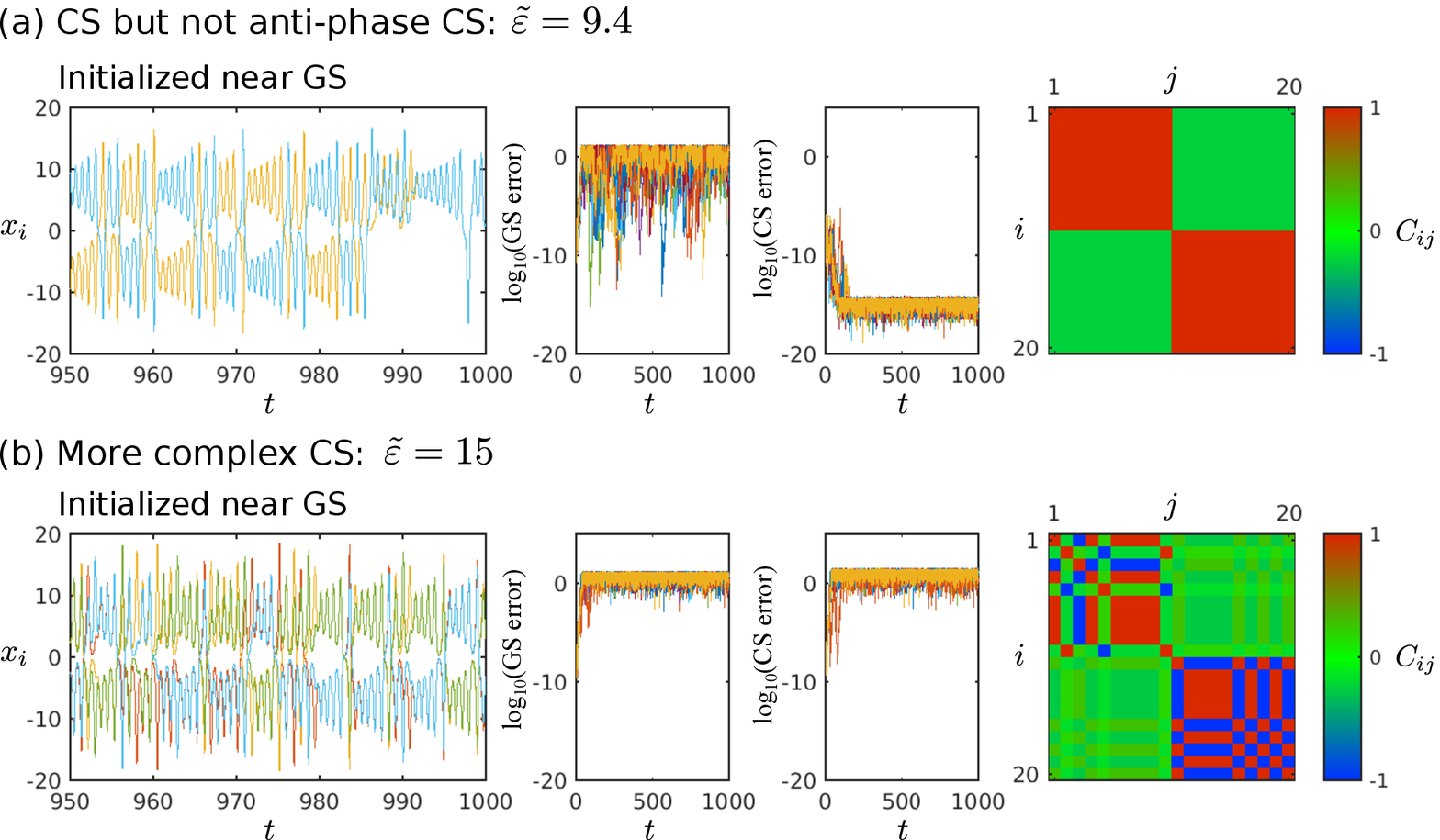}
\end{center}
\caption{\label{fig:direct-sim-lorenz-3}
Synchronization dynamics in UCM networks of coupled Lorenz oscillators with $\ell=2$ groups of size $k=10$ for $\tilde{\varepsilon}=9.4$ and $\tilde{\varepsilon}=15$. Detail description of individual plots is the same as in Fig.~\ref{fig:direct-sim-lorenz}.}
\end{figure*}

Although this linear stability transition scenario is clear and simple, the actual dynamics of the network is more subtle and rich.
Toward the end of the bi-stability regime $\tilde{\varepsilon}^\text{th}_{1} < \tilde{\varepsilon} < \tilde{\varepsilon}^\text{th}_2$, trajectories appear to hop back and forth between the GS and CS states, regardless of whether they are initialized near GS or CS state, which is reminiscent of riddled basins.\cite{Alexander:1992,Ott:1994}
This is illustrated in Fig.~\ref{fig:direct-sim-lorenz-2}(a), which shows an example trajectory initialized near a GS state that subsequently approach a CS state and then a GS state (top panels), as well as one initialized near a CS state that approach a GS state and then a CS state (bottom panels).
Evidence of riddled basins has been reported for coupled R\"{o}ssler systems in Ref.~\onlinecite{Pecora:2015}.
Beyond $\tilde{\varepsilon}^\text{th}_2$, we observe similar behavior, despite the absence of stable GS state (Fig.~\ref{fig:direct-sim-lorenz-2}(b), top panels).  This can be interpreted as chaotic intermittent bursts from the CS state.
After the CS state loses its stability at $\tilde{\varepsilon}^\text{th}_3$, we observe a CS state that is not quite the anti-phase CS state, in which the dynamics of different clusters are not perfectly anti-correlated (Fig.~\ref{fig:direct-sim-lorenz-3}(a)).
Increasing $\tilde{\varepsilon}$ further, we find a very different state (Fig.~\ref{fig:direct-sim-lorenz-3}(b)), in which neither of the two groups defined by the structure of the UCM network are synchronized.
The correlation matrix in Fig.~\ref{fig:direct-sim-lorenz-3}(b) shows a more complex pattern of synchrony, in which some oscillators are anti-phase synchronized ($C_{ij}\approx -1$) and others are not synchronized at all ($C_{ij}$ closer to $0$).

\section{Conclusions}

The symmetry of a system has profound impact on its dynamical behavior.\cite{Golubitsky:2003}
This is particularly insightful for networked systems, for which characterizing the relation between the network structure and dynamics is nontrivial.
Our results show that the significant role played by network symmetry for CS stability\cite{Pecora:2014zr} can be exploited to provide systematic analysis of systems with strong network-structural symmetry, such as the UCM networks.
Yet, our numerics suggests that bifurcations involving the GS and CS states have much more depth than is expected from the MSF-based stability analysis.
A complete characterization of the basins of attraction of these states in the high-dimensional phase space is generally a challenging problem, and our simulation results offer a glimpse of the rich structure it appears to have.
Another possible extension of our work is to consider a variety of other possible patterns of cluster synchronization, which can arise from properties of the system other than network symmetry, such as the internal symmetry of the node dynamics\cite{Benoit:1996} and the Laplacian property of the coupling matrix\cite{Sorrentino:2016}.

The fact that network optimization leads to highly symmetric structure in UCM networks suggests a general principle that 
optimization 
tends to generate symmetry in networks.
While noise, delays, and imperfections may prevent exact symmetries in real systems (or real implementation of such optimal networks), we expect an elevated level of symmetry in networks under pressure to optimize 
their 
function, and there is indeed evidence for abundance of symmetry in real networks.\cite{MacArthur:2008} 
We suggest that identifying mechanisms by which optimization induces symmetry structures will improve our understanding of the structure-dynamics relation in networks; moreover, it will help design systems with optimized functionality.

\begin{acknowledgments}
The authors thank J.~Lehnert and Y.~S.~Cho for insights on stability analysis for cluster synchronous states, as well as Y.~Zhang for valuable discussion on symmetry in synchronization-optimal networks.
This work was supported by ARO Grant No. W911NF-15-1-0272.
\end{acknowledgments}

%:References

\end{document}